*Original research*

# Pentamodes: the role of unit cell base topology on mechanical properties

Moein Shafia[1], Kaivan Mohammadi[1], Javad Akbari[1], Reza Hedayati[2,*]

[1]Advanced Manufacturing Lab (AML), Department of Mechanical Engineering, Sharif University of Technology, Azadi Avenue, 11365-11155 Tehran, Iran

[2]Aerospace Materials and Structures Department, Faculty of Aerospace Engineering, Delft University of Technology (TU Delft), Kluyverweg 1, 2629 HS Delft, the Netherlands

*Corresponding author. Email: r.hedayati@tudelft.nl; rezahedayati@gmail.com

## Abstract

Pentamodes (first conceived theoretically by Milton and Cherkaev) are a very interesting class of mechanical metamaterials where the bulk and shear moduli are decoupled. The pentamodes usually are composed of double cone-shaped struts with the middle diameter being large and the end diameters being tiny (ideally approaching zero). The cubic diamond geometry was proposed by Milton and Cherkaev as a suitable geometry for the unit cell and has since been used in the majority of the works on pentamodes. In this work, we aim to evaluate the degree to which the main unit cell design is contributing to high bulk to shear modulus ratio (known as FOM). In addition to the diamond unit cell, three other well-known unit cell designs are considered, and the effect of small diameter radius and the ratio of large-to-small diameter ratio, α, on the FOM is evaluated. The results showed that regardless of the primitive unit cell shape, the FOM value is highly dependent on the $d$ value, but its dependence on the $D$ value is very weak. For $d/h \propto 0.05$ ($h$ representing the linkage length), figures of merit in the range of $10^3$ could be reached for all the studied topologies.

# 1. Introduction

Mechanical metamaterials are a new class of designer materials that exhibit elastic properties rarely found in nature [1-3]. Such rationally-designed materials include materials with negative Poisson's ratio (known as auxetics) [4-6], materials with negative stiffness [7, 8], and pentamodes [9-11].

The inverse problem of finding a microstructure that can give the wanted mechanical properties is very difficult. This can be attributed to the fact that, unlike many other fields such as heat conduction, electric conduction, diffusion, etc., the underlying equations for mechanical response of continuum-mechanics equations are not form-invariant [12]. Direct lattice transformation approaches are simpler and easier to implement. Many efforts have been made toward design approaches to gain prescribed constitutive properties, the first of which can be attributed to the pentamode structures introduced theoretically by Milton and Cherkaev [13].

Pentamodes are metamaterials in which the bulk and shear moduli are decoupled. Such materials have very large bulk moduli compared to their shear moduli [14-16], leading to the usage of the term "metafluids" for them [17]. Ideally, the 6×6 elasticity tensor of these materials has 5 ("penta") zero members, and only one member is non-zero [13, 17]. The pentamodes usually are composed of double cone-shaped struts with the middle diameter being large and the end diameters being tiny (ideally approaching zero). A perfect pentamode metamaterial flows away with very small shear forces, and that's why, in practice, a small but "finite" shear modulus is needed. Therefore, in the majority of the previously manufactured pentamode metamaterials [14, 17], the end diameters of the struts are small (but non-zero). Using numerical and experimental studies, Kadic et al. [15], Martin et al. [11], Schittny et al. [14], and Hedayati et al. [18] have shown that the smaller diameter of the double cones (in the vertices of the lattice structure) are the most determinant parameter in the mechanical properties of low-density pentamode metamaterials.

Milton and Cherkaev [13] proposed a diamond unit cell for such a structure in which four linkages meet in equal angles at each vertex of the structure. They proposed this morphology in analogy with bimode metamaterials, which can only support a single stress in two-dimensional space. In their proposed design for bimode, three linkages met at a point. Therefore, they concluded that a natural way towards 3D extremal material (pentamode) is to consider a unit cell shape where only four linkages meet at the vertices. This diamond geometry has since been used in majority of the works on pentamodes.

In this work, we aim to evaluate the degree to which the main unit cell design contributes to high bulk-to-shear modulus ratio (known as FOM). In addition to the diamond unit cell, three other well-known unit cell designs are considered, and the effect of small diameter radius and the ratio of large-to-small diameter ratio, $\alpha$, on the FOM is evaluated.

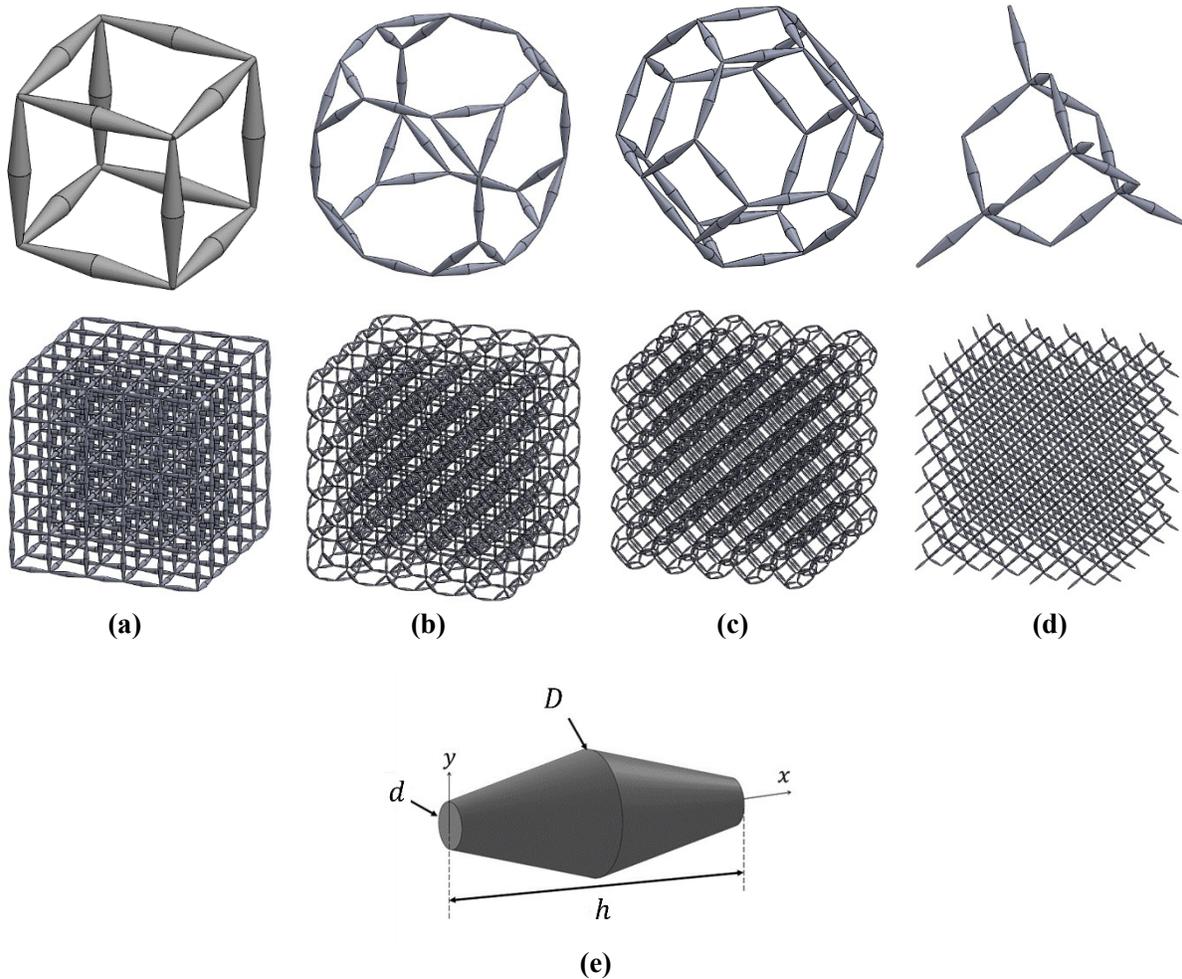

Figure 1: *The unit cells and corresponding lattice structures considered for this study: (a) cube, (b) truncated cube, (c) truncated octahedron, and (d) diamond topologies. (e) The dimensions of struts.*

## 2. Materials and methods

Creality LD_002R LCD 3D printer with ESUN PLA resins (Polyurethane acrylate) was used to manufacture all the specimens with a layer height of 30 μm. The curing time was set to 7 s. The geometrical dimensions of $h = 4\ mm$, $D = 1200\ \mu m$, and $d = 400\ \mu m$ were used for all the manufactured specimens. All the lattice structures were composed of 5×5×5 unit cells in each direction. The dimensions of the lattice structures based on cube, truncated cube, and truncated octahedron unit cells were $21.42 \pm 0.45\ mm$, $49.72 \pm 0.37\ mm$, and $57.83 \pm 0.15\ mm$, respectively. The mechanical properties of the constituent polymer were obtained using dog-bone specimens according to ISO D638-14. The bulk material had an elastic modulus of $E_s = 0.51\ GPa$, yield strength of $\sigma_{ys} = 28.0\ MPa$, Poisson's ratio of $\nu_s = 0.4$, and density of $\rho_s = 1020\ kg/m^3$.

The compression experimental tests were performed using SANTAM STM-20 (Tehran, Iran) mechanical test bench under the displacement rate of 2.5 mm/min. 20 kgf load cells were implemented to measure the load level in the truncated cube and truncated octahedron topologies, while 100 kgf load cells were used for measuring the loads in cube topology.

Based on the microscopic measurements performed on the manufactured specimens, numerical finite element (FE) models were constructed and solved in COMSOL Multiphysics package (Sweden). In the models constructed for uniaxial compressive loading, all the nodes at the lowermost surface of the lattice structures were allowed to move in their plane only, and they were constrained in the vertical direction. The upper surface of the lattice structure was displaced downwards for 1 mm. To obtain the elastic modulus, the elastic energy absorbed in the structure was measured, and inserted in the relationship $E = 2UL/A\delta^2$, where $U$, $L$, $A$, and $\delta$ denote the strain energy, dimension of the lattice structure parallel to the loading direction, structure's cross-sectional area perpendicular to the loading direction, and the resulting displacement. In the models constructed for hydrostatic compressive loading, in each cartesian direction, one face of the lattice structure was constrained, and the other face was displaced uniformly. The bulk modulus was obtained from the elastic energy: $B = 2U/9L\delta^2$. The shear modulus was obtained from $G = 2UL/A\delta^2$. Mesh sensitivity analyses was performed, and it was found that element sizes in the range of 14-110 $\mu m$ gives acceptable results.

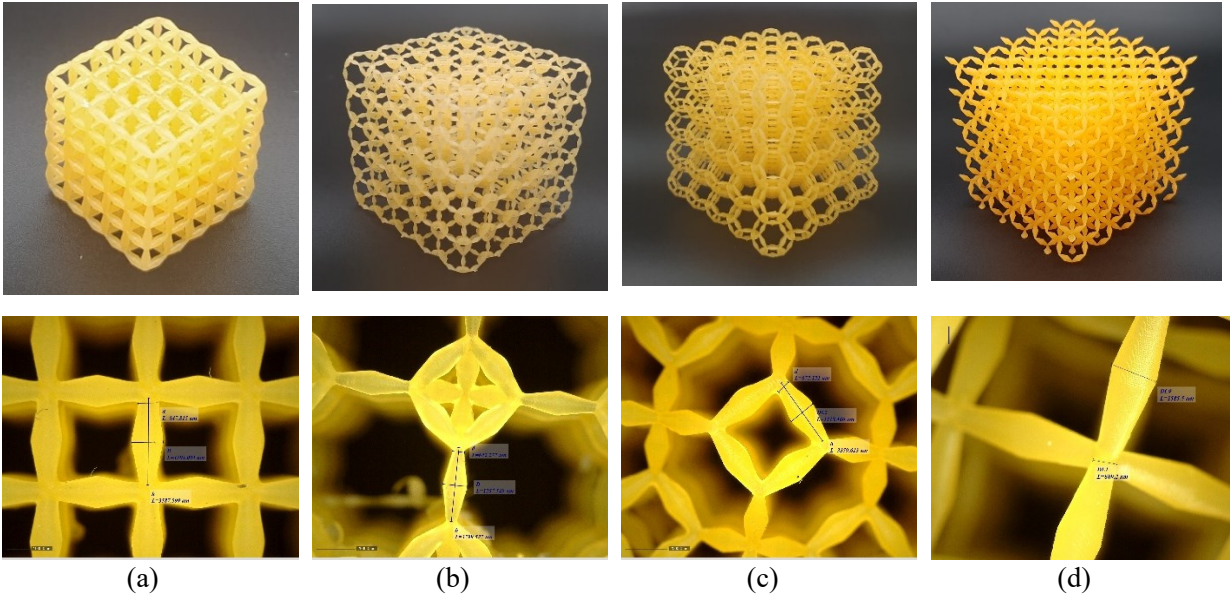

(a) (b) (c) (d)

*Figure 2: The macroscopic and microscopic images of the manufactured pentamodes based on (a) cube, (b) truncated cube, (c) truncated octahedron, and (d) diamond topologies.*

## 3. Results and discussions

The comparison of elastic modulus obtained from the numerical model and experimental tests was less than 9.5%, 21.5%, and 15.5% for the cubic, truncated cube, and truncated octahedron structures, respectively. The elastic modulus obtained for the cube structure was one order of magnitude greater than the corresponding values in other geometries. In general, increasing the smaller diameter $d$ increased the elastic modulus in all the geometries exponentially (Figure S2). The exponential effect of increasing $d$ on the elastic modulus was more significant in the diamond structure compared to other geometries. The relative elastic modulus for small diameter of $d = 400 \ \mu m$ and for $\alpha = 5$ was $3.9 \times 10^{-6}$, $5.27 \times 10^{-4}$, $6.7 \times 10^{-3}$, and $2.56 \times 10^{-6}$ for the truncated octahedron, truncated cube, cube, and diamond lattice structures (Figure S2). The effect of $\alpha$ on the increase in the elastic modulus was almost linear for the cubic structure, while it had an exponential effect on other topologies (Figure S2). As expected, the highest strain and stress levels occurred at the tip of double-cones in all the structures (Figure 3).

Increasing the smaller diameter had a monotonic decreasing effect on the Poisson's ratio of the truncated octahedron, while it had negligible effect on the Poisson's ratio of the truncated cube and diamond structures, and it had a non-monotonic effect on the cubic structure (Figure S3). For comparison purposes, for all the ranges of $d$ and $\alpha$, the Poisson's ratio of the truncated octahedron, truncated cube, cube, and diamond lattice structures were in the range of $0.49 - 0.57$, $0.15 - 0.22$, $0.04 - 0.4$, and $0.48 - 0.62$, respectively (Figure S3). Increasing the $\alpha$ value had a negligible effect on the Poisson's ratio of the truncated octahedron and diamond structures, while its effect on the the cube structure was non-monotonic (Figure S3). Increasing the $\alpha$ value changed the Poisson's ratio of the truncated cube structure only for very large $d$ and $\alpha$ (Figure S3d).

Increasing the smaller diameter $d$ had a monotonic increasing effect on the yield stress of the truncated cube and diamond structures (Figure S4). In the truncated octahedron and cube topologies, on the other hand, the exponential growth diminished at $d = 1000 \ \mu m$ (Figure S3). The relative yield stress for the small diameter of $d = 400 \ \mu m$ and for $\alpha = 5$ was $1.1 \times 10^{-6}$, $8.16 \times 10^{-5}$, $1.1 \times 10^{-3}$, and $2.98 \times 10^{-8}$ for the truncated octahedron, truncated cube, cube, and diamond lattice structures, respectively (Figure S4). Therefore, the yield strength of the cubic structure was three orders of magnitude greater than that of the truncated cuboctahedron and five orders of magnitude greater than diamond structure. Increasing the $\alpha$ ratio had a more or less increasing effect on the yield stress of all the structures other than truncated octahedron (Figure S4). In the truncated octahedron topology, the $\sigma_y - \alpha$ curve could get a bell-shaped geometry (Figure S4b).

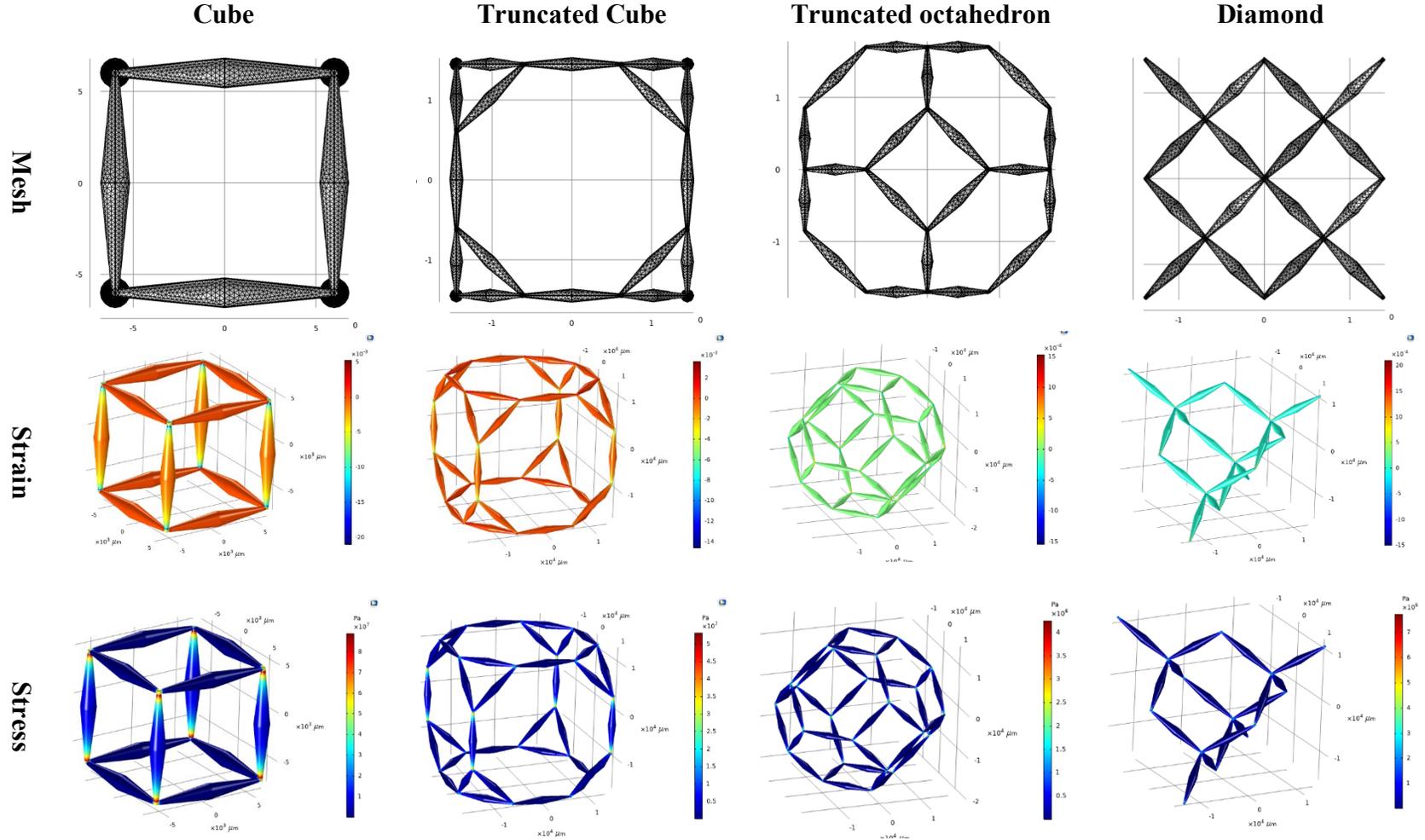

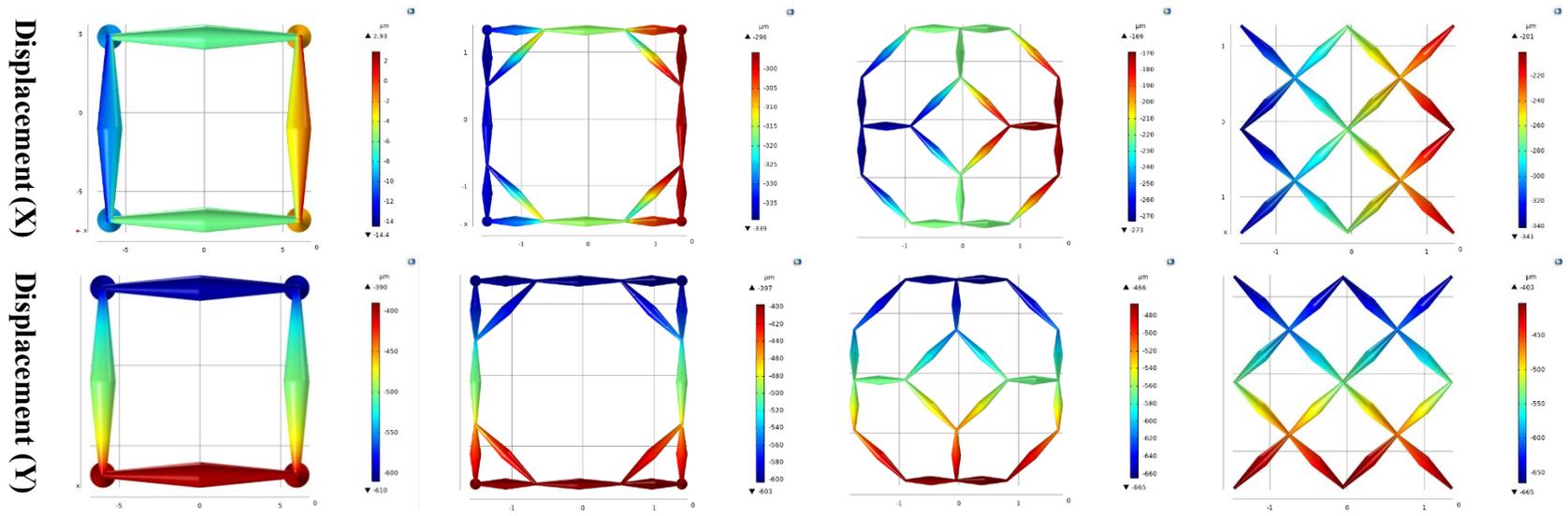

*Figure 3: The discretization, strain contour, stress contour, and displacements for pentamodes based on cube, truncated cube, truncated octahedon, and diamond topologies (d = 400 µm and α = 4)*

Increasing the smaller diameter $d$ had a monotonic decreasing effect on the $B/G$ ratio of the truncated octahedron and diamond structures (Figure 4). The $B/G$ ratio in the truncated cube topology had a peak at $d = 400~\mu m$ for $\alpha = 5$ and $\alpha = 4$ (Figure 4). The $B/G$ ratio for the small diameter of $d = 400~\mu m$ and for $\alpha = 5$ was 456, 575, 640, and 600 for the truncated octahedron, truncated cube, cube, and diamond lattice structures (Figure 4). The greatest measured $B/G$ ratio for the truncated octahedron, truncated cube, cube, and diamond lattice structures was 835, 1273, 3092, and 1880, respectively. This shows that the cubic structure had the greatest $B/G$ ratio, while the truncated octahedron demonstrated the smallest $B/G$ ratio. After the cubic structure, the greatest $B/G$ ratio belonged to the diamond structure.

Independence of $B/G$ from $\alpha$ ratio has always been an important characteristic of pentamode metamaterials. Hence, studying the effect of $\alpha$ on $B/G$ in topologies other than diamond is very important. The results show that the pentamode based on diamond unit cell is almost independent from $\alpha$ for $d = 400 - 1000~\mu m$ (Figure 4h). Some small dependencies of $B/G$ on $\alpha$ can be observed for the diamond structure with $d = 200~\mu m$ (Figure 4h). Interestingly, the $B/G$ ratio of the truncated octahedron and cube topologies are independent from $\alpha$ for all small diameter values (Figure 4b,f). The truncated cube structure is the only structure that shows some dependencies on the $\alpha$ ratio, although it is relatively small for $d \geq 600~\mu m$.

Comparing all the $B/G$ ratios at all ranges of $\alpha$ and small diameters $d$, in small $d$, i.e. $d = 200~\mu m$, the cubic structure has a very high $B/G$ ratio in the range of $3 \times 10^3$, while the truncated octahedron unit cell has the smallest $B/G$ ratio (Figure 5). In large $d$, i.e. $d = 1000~\mu m$, the truncated cube structure has the highest $B/G$ ratio (around 120), while the cubic unit cell has the smallest $B/G$ ratio (Figure 5).

## Truncated octahedron

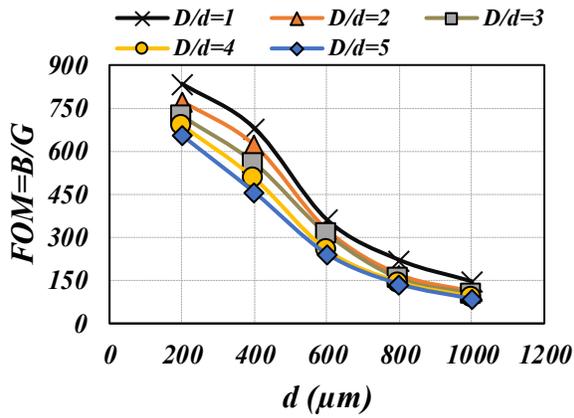
(a)

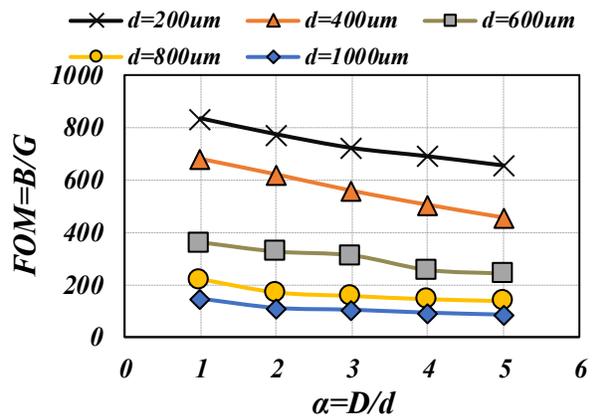
(b)

## Truncated cube

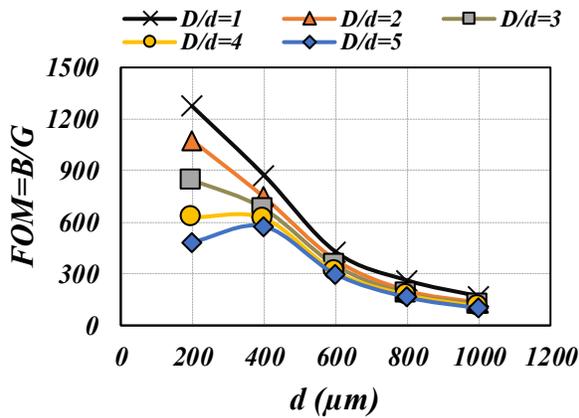
(c)

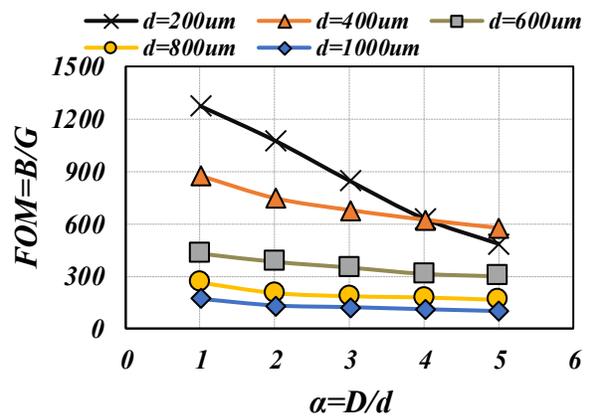
(d)

## Cube

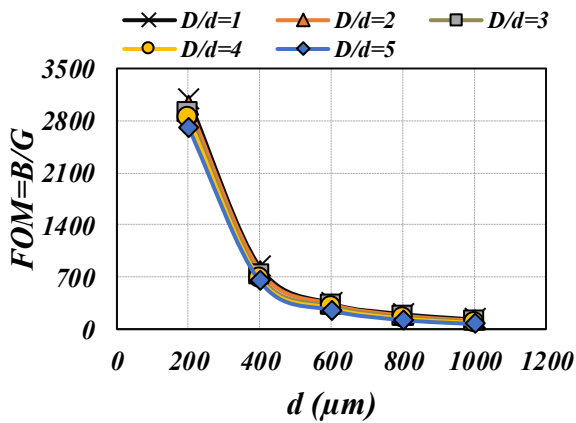
(e)

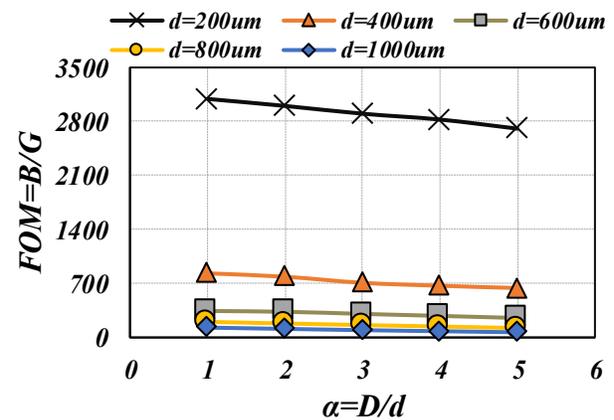
(f)

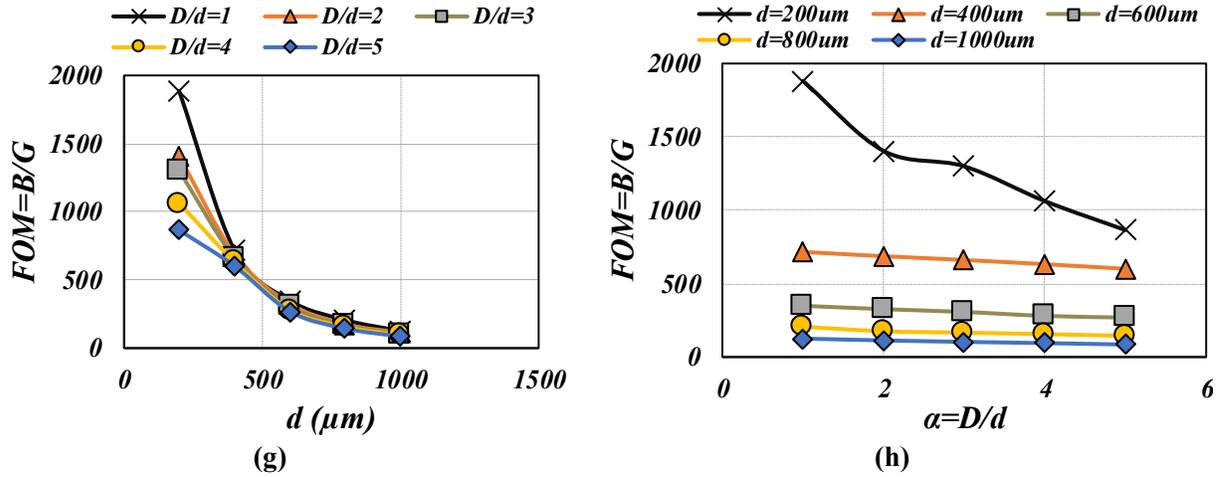

*Figure 4: Effect of variation of the connection point diameter d (left column) and α (right column) on the ratio of bulk modulus to shear modulus for pentamodes based on (a-b) truncated octahedon, (c-d) truncated cube, (e-f) cube, and (g-h) diamond topologies*

Looking at the results obtained, it was seen that for all the topologies, the $B/G$ ratio is almost independent from the $\alpha$ value. This is because after exerting an external load, the stress is mostly concentrated at the touching points of the double-cones, and the other parts do not significantly influence the loadbearing capacity of the structure. Therefore, by varying the size of the thick part, one can adjust the density of the lattice structure without changing its mechanical response over a large range [15]. This can lead to structures with decoupled modulus and relative density, in contrast to other porous materials [19-21] which show a strong correlation between their density and mechanical properties [20, 22, 23]. Such a property can be very appealing in designing biomedical implants where it is desired to have independent distributions of mechanical properties and permeability [18, 24].

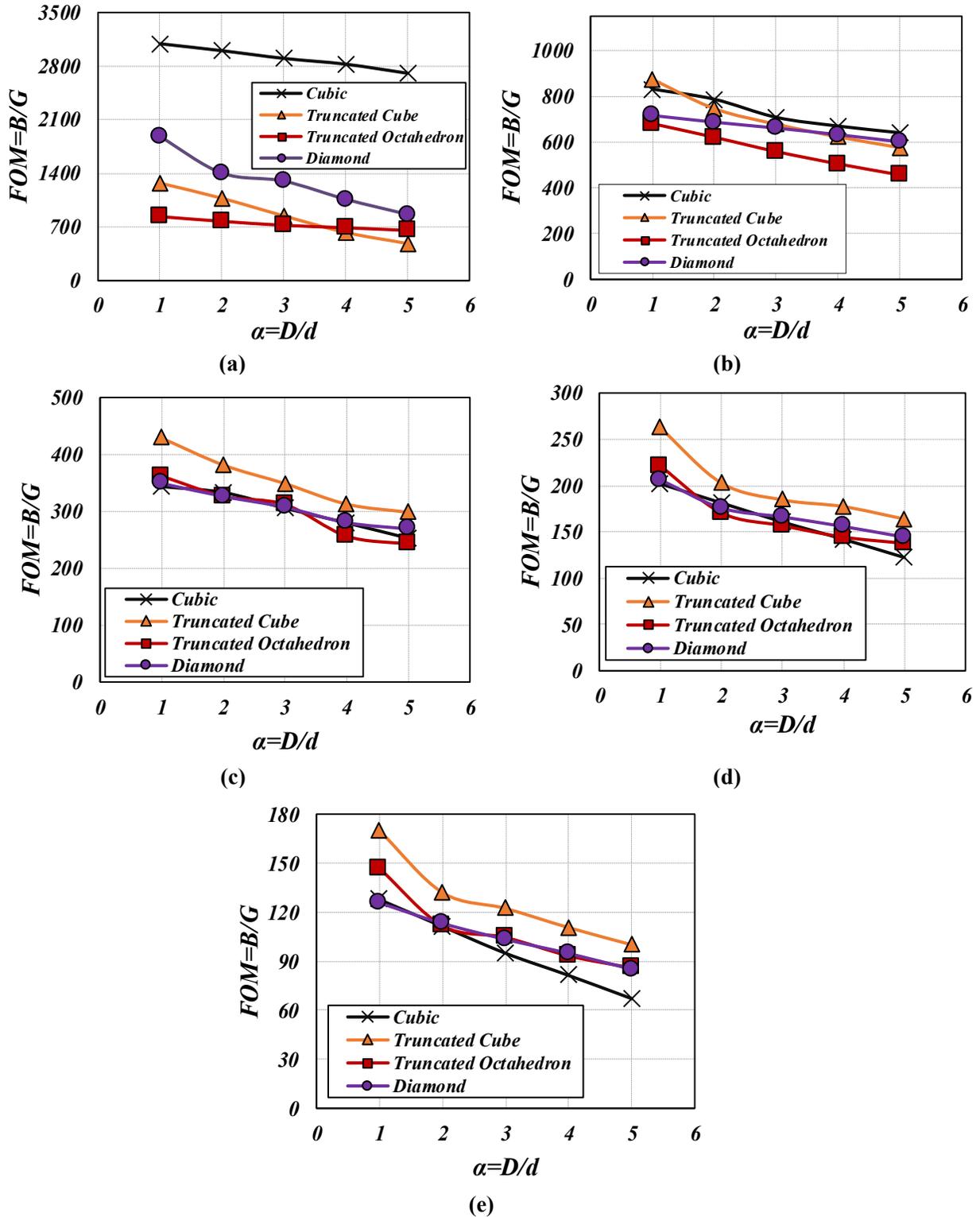

*Figure 5: Effect of variation of α on the ratio of bulk modulus to shear modulus for smaller diameters of (a) 200 μm, (b) 00 μm, (c) 600 μm, (d) 800 μm, and (e) 1000 μm*

## 5. Conclusions

In summary, the mechanical properties of four types of lattice structures (cube, truncated cube, diamond, and truncated octahedron) with double-cone struts were studied experimentally and numerically, and the effect of size of the smaller $d$ and larger $D$ diameters of the double-cones on the FOM or the ratio of $B/G$ was studied. It was observed that regardless of the primitive unit cell shape, the FOM value is highly dependent on the tip diameter $d$ value, but its dependence on the thicker part with $D$ value is very weak. For $d/h \propto 0.05$ ($h$ representing the linkage length), figures of merit in the range of $10^3$ could be reached for all the topologies considered.

## Data Availability Statement

The data that support the findings of this study are available from the corresponding author upon reasonable request.

## Supplementary Material

See Supplementary Material document for stress-strain curves of pentamode structures, variation of elastic modulus graphs, variation of Poisson's ratio graphs, variation of yield stress graphs, deformed images of the specimens, and displacement contour images.

## Conflicts of Interest

The authors have no conflicts to disclose.

## Authors Contributions

All authors contributed equally to this work.

## References

1. Wu, W., W. Hu, G. Qian, H. Liao, X. Xu, and F. Berto, *Mechanical design and multifunctional applications of chiral mechanical metamaterials: A review.* Materials & design, 2019. **180**: p. 107950.
2. Barchiesi, E., M. Spagnuolo, and L. Placidi, *Mechanical metamaterials: a state of the art.* Mathematics and Mechanics of Solids, 2019. **24**(1): p. 212-234.
3. Zhao, A., X. Zhang, W. Yu, Z. Zhao, X. Cai, and H. Chen, *Design and simulation of broadband multiphase pentamode metamaterials.* Applied Physics Letters, 2021. **118**(22): p. 224103.
4. Hedayati, R., A. Güven, and S. Van Der Zwaag, *3D gradient auxetic soft mechanical metamaterials fabricated by additive manufacturing.* Applied Physics Letters, 2021. **118**(14): p. 141904.
5. Yang, L., O. Harrysson, H. West, and D. Cormier, *Mechanical properties of 3D re-entrant honeycomb auxetic structures realized via additive manufacturing.* International Journal of Solids and Structures, 2015. **69**: p. 475-490.
6. Hedayati, R., N. Roudbarian, S. Tahmasiyan, and M. Bodaghi, *Gradient Origami Metamaterials for Programming Out-of-plane Curvatures.* Advanced Engineering Materials, 2023.

7. Hewage, T.A., K.L. Alderson, A. Alderson, and F. Scarpa, *Double-Negative Mechanical Metamaterials Displaying Simultaneous Negative Stiffness and Negative Poisson's Ratio Properties.* Advanced Materials, 2016. **28**(46): p. 10323-10332.
8. Sun, Y. and N.M. Pugno, *In plane stiffness of multifunctional hierarchical honeycombs with negative Poisson's ratio sub-structures.* Composite Structures, 2013. **106**: p. 681-689.
9. Mohammadi, K., M. Movahhedy, I. Shishkovsky, and R. Hedayati, *Hybrid anisotropic pentamode mechanical metamaterial produced by additive manufacturing technique* Applied Physics Letters, 2020. **117**(6): p. 061901.
10. Krushynska, A., P. Galich, F. Bosia, N. Pugno, and S. Rudykh, *Hybrid metamaterials combining pentamode lattices and phononic plates.* Applied Physics Letters, 2018. **113**(20): p. 201901.
11. Martin, A., M. Kadic, R. Schittny, T. Bückmann, and M. Wegener, *Phonon band structures of three-dimensional pentamode metamaterials.* Physical Review B, 2012. **86**(15): p. 155116.
12. Bückmann, T., M. Kadic, R. Schittny, and M. Wegener, *Mechanical cloak design by direct lattice transformation.* Proceedings of the National Academy of Sciences, 2015. **112**(16): p. 4930-4934.
13. Milton, G.W. and A.V. Cherkaev, *Which elasticity tensors are realizable?* Journal of engineering materials and technology, 1995. **117**(4): p. 483-493.
14. Schittny, R., T. Bückmann, M. Kadic, and M. Wegener, *Elastic measurements on macroscopic three-dimensional pentamode metamaterials.* Applied Physics Letters, 2013. **103**(23): p. 231905.
15. Kadic, M., T. Bückmann, R. Schittny, P. Gumbsch, and M. Wegener, *Pentamode metamaterials with independently tailored bulk modulus and mass density.* Physical Review Applied, 2014. **2**(5): p. 054007.
16. Huang, Y., X. Lu, G. Liang, and Z. Xu, *Pentamodal property and acoustic band gaps of pentamode metamaterials with different cross-section shapes.* Physics Letters A, 2016. **380**(13): p. 1334-1338.
17. Kadic, M., T. Bückmann, N. Stenger, M. Thiel, and M. Wegener, *On the practicability of pentamode mechanical metamaterials.* Applied Physics Letters, 2012. **100**(19): p. 191901.
18. Hedayati, R., A. Leeflang, and A. Zadpoor, *Additively manufactured metallic pentamode meta-materials.* Applied Physics Letters, 2017. **110**(9): p. 091905.
19. Ghavidelnia, N., R. Hedayati, M. Sadighi, and M. Mohammadi-Aghdam, *Development of porous implants with non-uniform mechanical properties distribution based on CT images.* Applied Mathematical Modelling, 2020. **83**: p. 801-823.
20. Hedayati, R., N. Ghavidelnia, M. Sadighi, and M. Bodaghi, *Improving the accuracy of analytical relationships for mechanical properties of permeable metamaterials.* Applied Sciences, 2021. **11**(3): p. 1332.
21. Kelly, C.N., A.T. Miller, S.J. Hollister, R.E. Guldberg, and K. Gall, *Design and structure–function characterization of 3D printed synthetic porous biomaterials for tissue engineering.* Advanced healthcare materials, 2018. **7**(7): p. 1701095.
22. Zheng, X., H. Lee, T.H. Weisgraber, M. Shusteff, J. DeOtte, E.B. Duoss, J.D. Kuntz, M.M. Biener, Q. Ge, and J.A. Jackson, *Ultralight, ultrastiff mechanical metamaterials.* Science, 2014. **344**(6190): p. 1373-1377.
23. Ashby, M.F. and D. Cebon, *Materials selection in mechanical design.* Le Journal de Physique IV, 1993. **3**(C7): p. C7-1-C7-9.
24. Zadpoor, A.A., *Mechanical meta-materials.* Materials Horizons, 2016.